\documentclass[12pt,preprint]{aastex}
\usepackage{amsmath}
\usepackage{amsfonts}
\usepackage{amssymb}
\usepackage{xcolor}
\usepackage{enumerate}
\usepackage{color,soul}
\usepackage{makerobust}

\newcommand{\Mpch}{\mathinner{h^{-1} \mathrm{Mpc}}}

\newcommand{\kpch}{\mathinner{h^{-1} \mathrm{kpc}}}
\newcommand{\kmsMpc}{\mathinner{\mathrm{km~s^{-1}Mpc^{-1}}}}

\newcommand{\Msunh}{\mathinner{h^{-1} {M_\odot}}}

\newcommand{\fref}[1]{Figure~\ref{fig:#1}}

\newcommand{\tref}[1]{Table~\ref{tab:#1}}
\newcommand{\sref}[1]{Section~\ref{sec:#1}}
\newcommand{\eref}[1]{Equation~(\ref{eq:#1})}

\renewcommand{\vec}[1]{\boldsymbol{#1}}

\def\LC93{\textsf{LC93}}
\def\B08{\textsf{B08}}
\def\J08{\textsf{J08}}
\def\M12{\textsf{M12}}
\def\V13{\textsf{V13}}
\def\HR4{\textsf{HR4}}
\newcommand{\PSB}{\textsf{PSB}}

\begin{document}

\author{Sungwook E. Hong, Changbom Park}
\affil{School of Physics, Korea Institute for Advanced Study}
\affil{85 Hoegiro, Dongdaemun-gu, Seoul 02455, Korea}
\email{swhong@kias.re.kr, cbp@kias.re.kr}
\and 
\author{Juhan Kim\footnote{Corresponding author}}
\affil{Center for Advanced Computation, Korea Institute for Advanced Study}
\affil{85 Hoegiro, Dongdaemun-gu, Seoul 02455, Korea}
\email{kjhan@kias.re.kr}

\title{The Most Bound Halo Particle-Galaxy Correspondence Model: Comparison between Models with Different Merger Timescales}

\begin{abstract}
We develop a galaxy assignment scheme that populates dark matter halos with galaxies by tracing the most bound member particles (MBPs) of simulated halos.
Several merger-timescale models based on analytic calculations and numerical simulations are adopted as the survival time of mock satellite galaxies.
We build mock galaxy samples from halo merger data of the Horizon Run 4 $N$-body simulation from $z = 12$--0.
We compare group properties and two-point correlation functions (2pCFs) of mock galaxies with those of volume-limited SDSS galaxies, with $r$-band absolute magnitudes of $\mathcal{M}_r - 5 \log h < -21$ and $-20$ at $z=0$. 
It is found that the MBP-galaxy correspondence scheme reproduces the observed population of SDSS galaxies in massive galaxy groups ($M > 10^{14} \Msunh$) and the small-scale 2pCF ($r_{\rm p} < 10 \Mpch$) quite well for the majority of the merger timescale models adopted.
The new scheme outperforms the previous subhalo-galaxy correspondence scheme by more than $2\sigma$.
\end{abstract}


\maketitle

\section{Introduction}\label{sec:intro}

In the standard $\Lambda$CDM cosmology, dark matter halos grow hierarchically over cosmic time through mergers of smaller halos.
Dark matter halos provide a cradle site for stars and galaxies \citep{white1978, fall1980, blumenthal1984}.
Even after formation, galaxies are believed to be strongly influenced by the hierarchical clustering of their host halos.
Merging and accretion can trigger or regulate star formation, radiative cooling, supernova feedback, and chemical enrichment in galaxies. 
Frequent merger events of dark matter halos may leave fossilized evidence of the star formation history of galaxies.
Therefore, if one has detailed information on the mass history of a halo, one can better understand the internal properties and evolutions of the galaxies associated with the halo.

Among the various tools for studying galaxy formation, the hydrodynamical simulation describes the evolution of the baryonic content of galaxies by using the gas dynamics as well as gravity and astrophysical processes \citep{hernquist1989, monaghan1992, harten1997, tasker2006, weinberg2008}.
However, to include sub-galactic hydrodynamic processes in a volume sufficiently large enough to reduce cosmic variance, simulations should cover wide ranges of mass and length scales, which require  excessive computational power and complex parallel computing techniques.
Moreover, it requires details on baryonic physics, such as star formation, radiative cooling, supernovae feedback, and initial stellar mass function, which are not well-known.

The semi-analytic model (SAM), on the other hand, places galaxies at the position of the most bound member particles (MBPs) of simulated halos and subhalos.
To determine the various properties of each mock galaxy (e.g., luminosity, color, and star formation activity), the SAM applies analytic  prescriptions applied to the numerically found merging histories of host halos. 
However, the prescription parameters vary throughout literature, depending on which sets of prescriptions and observables are adopted to tune the parameters \citep{cole1994, kauffmann1997, springel2001, delucia2004, kang2005, baugh2006, merson2013}.
Furthermore, the semi-analytic recipes become more complicated as the number of target observables increases.

Alternatively, two types of approaches have been proposed to populate simulated halo galaxies while neglecting the details of the physics of galaxy formation and evolution.
The halo occupation distribution does not try to identify and characterize individual mock galaxies in simulations.
Instead, it aims to find the conditional probabilities of various galactic properties (e.g., luminosity, spatial distribution, and number density of satellites) for a given halo mass \citep{seljak2000, berlind2002, zheng2005}.
The conditional probabilities could be measured from observations \citep{abazajian2005, zheng2007, zehavi2011} or from numerical simulations \citep{jing1998, berlind2002, kravtsov2004}.

The subhalo-galaxy correspondence model assumes that each subhalo hosts only one galaxy \citep{conroy2006, kim2008, moster2010}. 
Physical properties of galaxies can be assigned to subhalos if one assumes a monotonic relationship between the subhalo and galaxy properties such as mass/luminosity or size.
However, it has been reported that such monotonic relationship may not properly work in cluster regions, where subhalos suffer from more tidal disruption than their embedded galaxies \citep{hayashi2003, kravtsov2004}.

Recent studies have claimed that one can avoid the above problem by adopting the infalling mass of a dark matter subhalo rather than its ongoing mass \citep{nagai2005, conroy2006, vale2006, conroy2009, moster2010}.
To know the infalling mass, one should have the mass history of a subhalo, which encounters the following issues:
(1) Because the (sub)halo merger history may be measured differently for different merger tree implementations, the physical properties of simulated galaxies may not be unique \citep{lee2014}.
(2) Most merger models adopt sophisticated merger-identification algorithms that require the history of all member particles.
Therefore, it may be difficult to apply those models to massive cosmological simulations with more than billions of subhalos.
(3) Subhalo finding may not work in a low simulation resolution.
Also, the infalling mass may vary depending on the subhalo-finding algorithms.

In this paper, we introduce an MBP-galaxy correspondence scheme that applies the modeled merger timescales to the fate of MBPs.
We develop our method with the following motivations:
(1) MBPs have been widely used in SAMs as one of the most reasonable proxies for galaxies \citep{delucia2004,faltenbacher2006}.
(2) MBPs enable us to build merger trees in a simple and computationally cheap way.
(3) Unlike traditional methods, our method can find  ``orphan galaxies,'' which have survived even after their host subhalos are disrupted \citep{gao2004}.
(4) Our method can be useful even in $N$-body simulations with relatively low  spatial and mass resolutions, where subhalo findings tend to have poor performances.

This paper is organized as follows.
In \sref{numerics}, we describe the details of our $N$-body simulation data, one-to-one correspondence model, and several models of the merger timescale.
In \sref{results}, we study the properties of our mock galaxy samples, such as the survival probability of satellite galaxies, galaxy group properties, and two-point correlation functions (2pCFs).
We also compare the properties of our mock galaxies with those of SDSS galaxies.
We summarize our results in \sref{summary}.

\section{Data and Models}\label{sec:numerics}
\subsection{Simulation Data}\label{sec:sim}

\begin{deluxetable}{lll}
\tablewidth{0pt}
\tablecaption{Horizon Run 4 Simulation Parameters\label{tab:sim_param}}
\tablehead{\colhead{Parameter} & \colhead{Value} & \colhead{Note}}
\startdata
$N_{\rm p}$ & $6300^3$ & Number of simulated particles\\
$L_{\rm box}$ & $3150\Mpch$ & Simulation box size in a length\\
$N_{\rm step}$ & 2001 & Number of time steps \\
$z_i$ & 100 & Initial redshift\\
$h$ & 0.72 & Hubble parameter in units of $100 \kmsMpc$ \\
$\Omega_{\rm m}$ & 0.26  & Matter density parameter\\
$\Omega_{\rm b}$ & 0.044 & Baryon density parameter\\
$\Omega_\Lambda$ & 0.74 & Cosmological constant parameter\\
$n_{\rm s}$ & 0.96 & Spectral index of power spectrum \\ 
$\sigma_8$ & 0.79 & RMS density fluctuation on the scale $R = 8 \Mpch$\\
$m_{\rm p}$ & $9.0 \times 10^9 \Msunh$ & Particle mass\\
$\epsilon_{\rm f}$ & $50 \kpch$ & Force resolution
\enddata
\end{deluxetable}

We use a cosmological $N$-body simulation called Horizon Run 4 (\HR4 hereafter), the latest one in our series of massive simulations \citep[][see \tref{sim_param}]{kim2015}.
While it has a comparable number of particles ($N_{\rm p} = 6300^3$), the \HR4 has 8--30 times smaller volume ($3150^3\,h^{-3} \mathrm{Mpc^3}$) than the previous Horizon Runs, so it allows us to study satellite halos in clusters. 

The \HR4 was performed by adopting the cosmological parameters of the Wilkinson Microwave Anisotropy Probe 5 year concordance $\Lambda$CDM cosmology \citep{dunkley2009}.
The initial displacement of each particle is calculated according to the second-order linear perturbation theory at the initial redshift ($z_{\rm i} = 100$) so that the initial displacement does not exceed the mean particle separation $d_{\rm mean} = 0.5 \Mpch$.
Then, the gravitational evolution of particles from $z_{\rm i}$ to $z_{\rm f}=0$ is calculated with the GOTPM \citep{dubinski2004} taking 2000 time steps so that the spatial shift of any particle in a given time step does not exceed the force resolution $\epsilon_{\rm f} = 0.1\,d_{\rm mean}$.

We identify halos from simulation data by using the friend-of-friend (FoF) method with a common choice of linking length $l_{\rm mFoF} = 0.2 \,d_{\rm mean}$.
The minimum number of member particles of a halo is set to 30, which leads to a minimum halo mass of $M_{\rm mFoF}^{\rm min} = 2.7 \times 10^{11} \Msunh$.

We trace merger trajectories of identified halos at 75 time steps between $z=12$ and 0, with a step size nearly equal to the dynamical timescale of the Milky-Way-sized galaxy.
A halo (hereafter $A$) is called an ancestor to another halo (hereafter $B$) found in the next time step if the MBP of $A$ is a member particle of $B$.
When a merger occurs, we call the MBP of the largest ancestor the host MBP of the merger remnant, while we called the other satellites MBPs.
We tag host and satellite MBPs with the ongoing masses and the infalling masses of their host halos, respectively.
We monitor the evolution of all MBPs inside halos until $z = 0$.

It should be noted that our merger tree of MBPs does not directly contain proper information on tidal disruptions.
Although we may implement the concept of orphan galaxies on the merger trees, we may need a model on the lifetime of satellites in a cluster environment.
Also, our merger tree does not distinguish merger or accretion from the fly-by interactions of halos.

\subsection{How to Link Galaxies to Halos}\label{sec:methods}

\begin{figure}[tpb]
\plotone{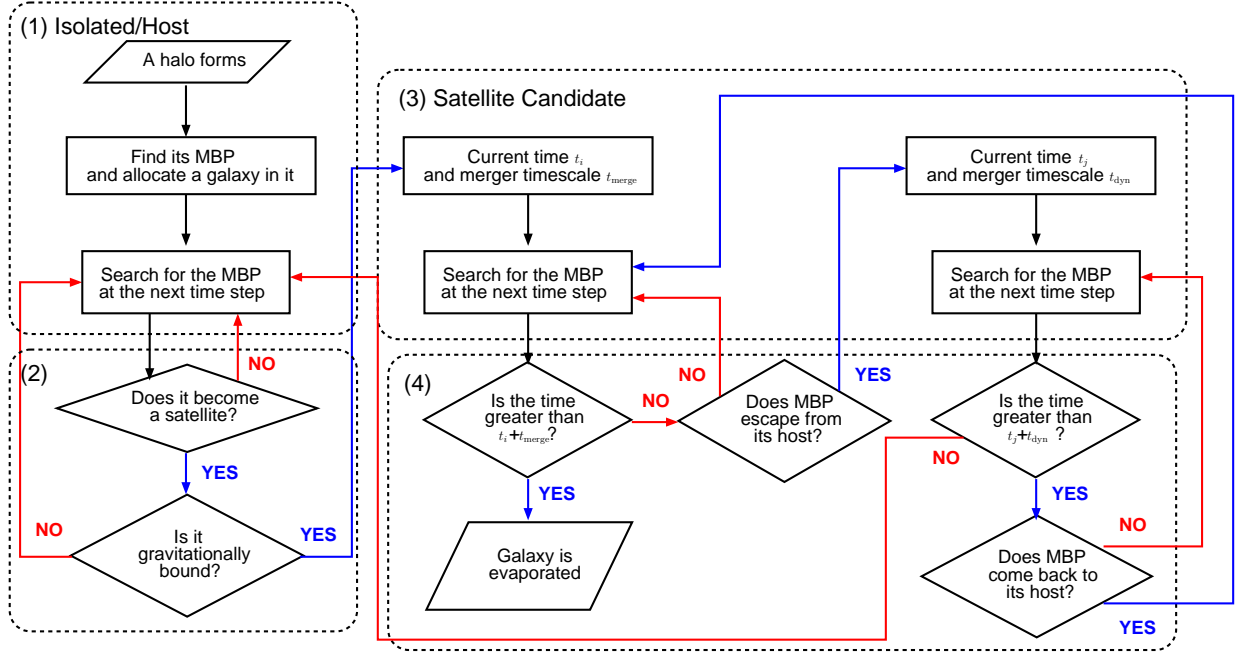}
\caption{Flowchart of galaxy assignment and satellite survival time decision.
(1) A galaxy is assigned to the MBP of an isolated halo. (2) A satellite galaxy is identified.
(3) The merger timescale is evaluated to determine the lifetime of a satellite. 
(4) If a satellite escapes its host halo before the tidal destruction, we set the satellite as an isolated galaxy again.}\label{fig:scheme}
\end{figure}

\fref{scheme} depicts the MBP-galaxy correspondence scheme applied to the \HR4 simulation data.
We mark all MBPs recorded in merger trees as galaxy proxies.
The mass, position, and velocity tagged to each MBP are used to model the galaxy luminosity, position, and velocity, respectively.
Luminosity is assigned to each mock galaxy using the abundance matching between the mass function of mock galaxies and the luminosity function of the SDSS main galaxies \citep{choi2007}.
However, because of the limitation of our merger trees described in the previous section, we need to calculate the survival time of a satellite galaxy. 
Also, we need to distinguish between an actual merger event and a fly-by event in a reasonable way.

Since galaxies are compact and gravitationally bound, one could expect that satellites would survive until they reach the centers of their host halos, where satellites merge into the central galaxy. 
Therefore, we define the survival time of a satellite galaxy as the merger timescale of its host halo ($t_{\rm merge}$). 
After complete tidal disruption, we terminate the tree link of the MBP.
In this paper, we test several theoretical models of the merger timescale proposed in the literature.

To differentiate between fly-bys and mergers, we use the following two-step process.
First, when we identify a merger candidate, we check whether a satellite MBP is gravitationally bound to its host.
If the total energy is positive, namely, if the satellite is not  bound, this merger candidate is dropped as a fly-by.
Once a satellite is gravitationally bound to a halo at $t_i$, we check whether it remains as the satellite until $t_i + t_{\rm merge}$.
If a satellite escapes before the estimated merger timescale, we check whether the escape is temporary or permanent.
To check it we use a dynamical timescale, 
\begin{equation}
t_{\rm dyn} \equiv \frac{R_{\rm vir}}{V_{\rm vir}} = \left( \frac{\Delta_{\rm vir}(z) H^2 (z)}{2} \right)^{-1/2},
\end{equation}
which is an orbital period around an object with a radius ($R_{\rm vir}$) and circular velocity \citep[$V_{\rm vir}$;][]{eke1996, bryan1998}.
Here, $H(z)$ is the Hubble parameter at redshift $z$, and $\Delta_{\rm vir}(z)$ is the mean density of a virialized object in a unit of the critical density at redshift $z$.
If an escaped satellite returns to its host within a dynamical timescale, we consider the escape being incidental.
If not, we mark the satellite as completely detached.

Due to the hierarchical clustering in the $\Lambda$CDM cosmology, a host halo with (a) satellite(s) may become a satellite to a bigger halo.
In this case, a single satellite MBP would have multiple host halos through its merger history, and it may be unclear which host halo should be applied to measure the merger timescale of the satellite.
In this paper, we assume that a satellite might be more affected by its closest host halo, or, the host halo of its earliest merger event.
For this reason, our reference model uses the host halo of the earliest merger event of a satellite to calculate $t_{\rm merge}$.
We also tried another model that uses the minimum merger timescale updated at every time step and compared the results with those of the reference model, though we found no significant statistical difference between the results of the two models.

\begin{figure}[tpb]
\plotone{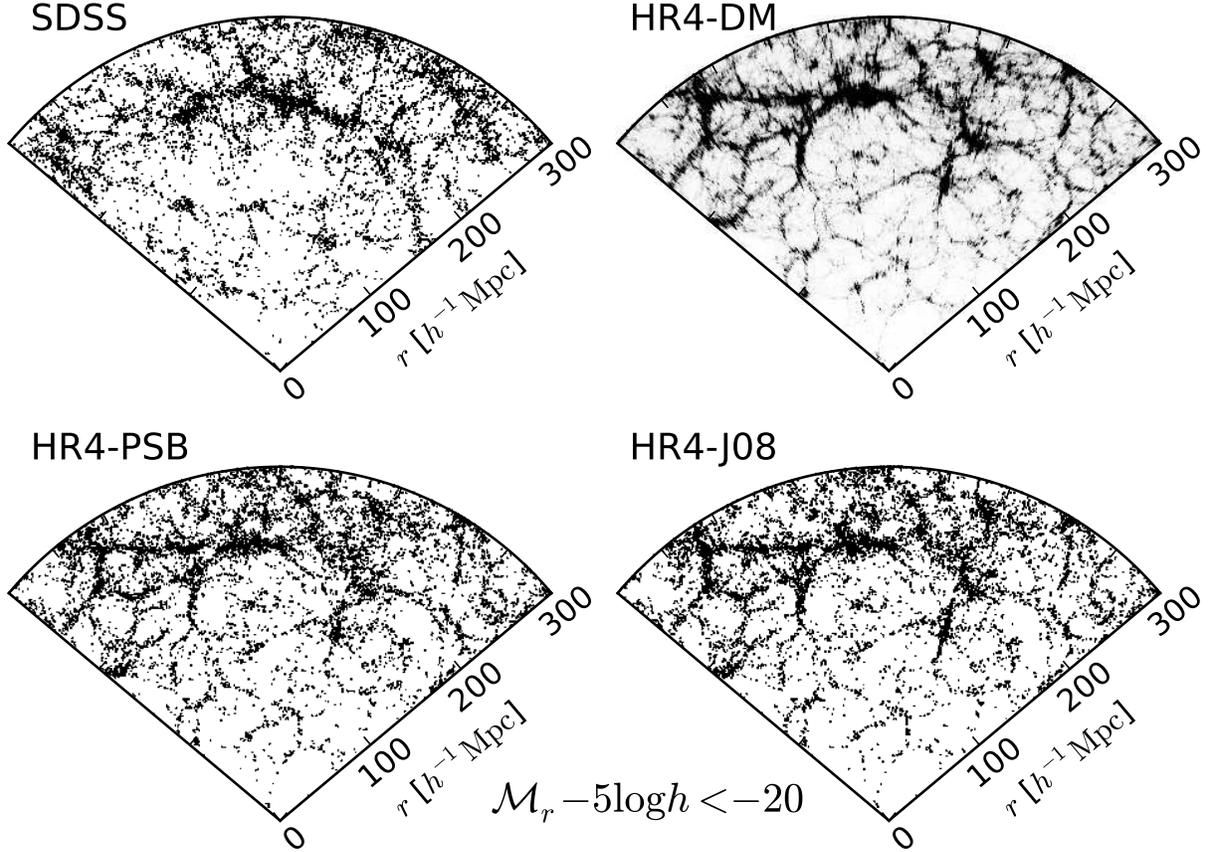}
\caption{Past-lightcone-space distributions of SDSS galaxies (top left) and simulations (other panels) at $z = 0$.
Top left: galaxies from the Korea Institute for Advanced Study Value-added Galaxy Catalog \citep[KIAS-VAGC;][]{choi2010a} with an $r$-band absolute magnitude of $\mathcal{M}_r - 5 \log h < -20$.
We show galaxies located in a survey region of the Survey coordinates $-33^{\circ} < \eta < -27^{\circ}$, $-50^{\circ} < \lambda < 50^{\circ}$.
The Sloan Great Wall is located at $r\simeq 200 ~h^{-1}{\rm Mpc}$, stretching along the tangential direction.
Top right: a matter density map of the \HR4 including redshift space distortion.
Bottom left: mock galaxies from the subhalo-galaxy correspondence scheme \citep{kim2008}.
Bottom right: same as in the bottom left, but from the MBP-galaxy correspondence scheme by applying the merger timescale model described in \cite{jiang2008}.}\label{fig:wedge}
\end{figure}

\fref{wedge} shows the spatial distribution of the volume-limited mock galaxies from the \HR4 with an $r$-band absolute magnitude of $\mathcal{M}_r - 5 \log h < -20$ in redshift space at $z=0$ (bottom right).
The corresponding galaxy number density is $\bar{n} \simeq 5.1 \times 10^{-3} h^3{\rm Mpc^{-3}}$.
From our visual inspection, the overall spatial distribution of our mock galaxies is similar to the SDSS galaxies with the same magnitude limit \citep[top left;][]{choi2010a}.
 
For a comparison between our MBP-galaxy and the traditional subhalo-galaxy correspondence schemes, 
we build mock galaxies from physically self-bound (PSB) subhalos \citep[bottom left;][]{kim2008}.
Here, the minimum number of member particles of a subhalo is set to 30, and the minimum subhalo mass is $M_{\rm PSB}^{\rm min} = 2.7 \times 10^{11} \Msunh$.
The mass, center of mass, and bulk velocity of a subhalo are used to determine the galaxy luminosity, position, and velocity, respectively.
In our previous studies, mock galaxies built from PSB subhalos reproduce several features of the observed galaxy distribution, such as the topology \citep{choi2010b, choi2013, parihar2014}, the largest-scale structure distribution \citep{park2012}, and the spin parameter distribution \citep{cervantes-sodi2008}.

\subsection{Models on Merger Timescale}\label{sec:models}

\begin{figure}[tpb]
\plotone{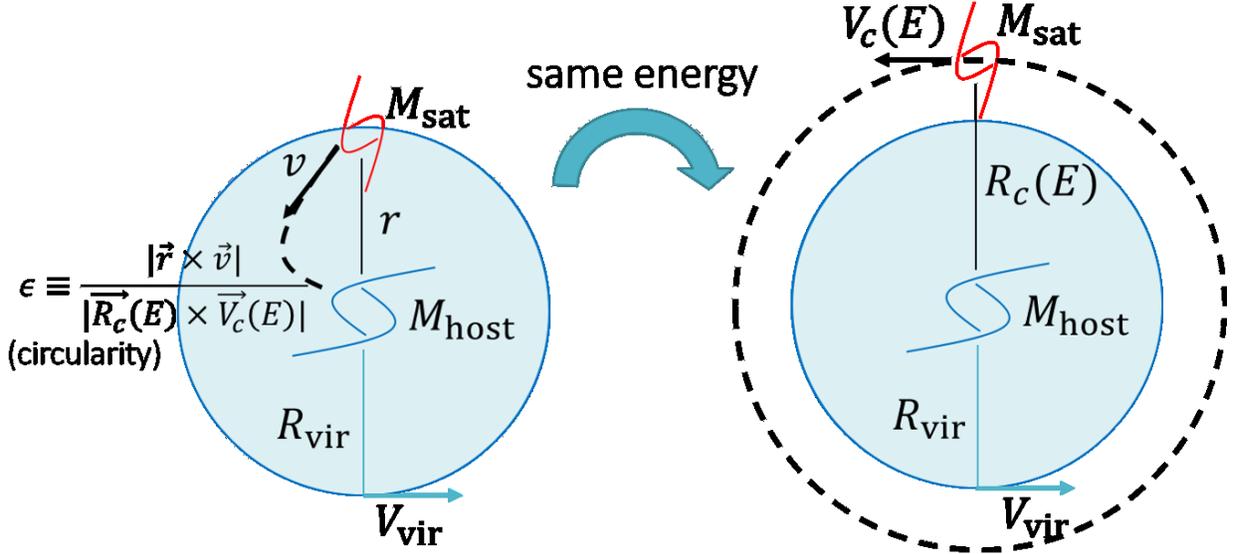}
\caption{Cartoon depicting the parameters used to estimate the merger timescale of a satellite. 
$\vec{r},~\vec{v}$: relative position and velocity of the satellite from its host.
$R_{\rm vir},~V_{\rm vir}$: Virial radius and the circular velocity at the virial radius of the host halo.
$M_{\rm sat}, ~M_{\rm host}$: masses of the satellite and its host.
$R_{\rm c}(E), ~V_{\rm c}(E)$: radius and velocity of an imaginary circular orbit of the satellite having an identical total energy.
$\epsilon$: circularity of the satellite orbit.
}\label{fig:merger_time_calculation}
\end{figure}

\begin{deluxetable}{llllll}
\tablewidth{0pt}
\tablecaption{Description of Merger Timescale Models\label{tab:model}}
\tablehead{
\colhead{Model} & \colhead{$f(\epsilon)$\tablenotemark{a}} & \colhead{$\alpha$\tablenotemark{a}} & \colhead{$\beta$\tablenotemark{a}} & \colhead{Method} & \colhead{Reference}}
\startdata
\LC93 & ${\epsilon^{0.78}}/{0.86}$ & 1 & 2 & Analytic & \cite{lacey1993} \\
\B08 & $0.216 \exp(1.9 \epsilon)$ & 1.3 & 1 & Isolated\tablenotemark{b}, $N$-body & \cite{boylan-kolchin2008} \\
\J08 & $(0.94 \epsilon^{0.60}+0.60)/0.86$ & 1 & 0 & Cosmo\tablenotemark{c}, SPH & \cite{jiang2008} \\
\M12 & $0.9 \exp(0.6\epsilon)$ & 1 & 0.1 & Cosmo\tablenotemark{c}, $N$-body & \cite{mccavana2012}\\
\V13 & $0.216 \exp(1.9 \epsilon) (1+z)^{0.44}$ & 1.3 & 1 & Isolated\tablenotemark{b}, SPH & \cite{villalobos2013} \\
\enddata
\tablenotetext{a}{Parameters defined in \eref{tmerge_general}}
\tablenotetext{b}{Isolated boundary condition}
\tablenotetext{c}{Periodic boundary condition}
\end{deluxetable}

\tref{model} is a summary of five adopted models for the merger timescale.
$t_{\rm merge}$ in these models share their functional forms with the analytic solution to an ideal case \citep{chandrasekhar1943, binney1987, lacey1993}: 
\begin{equation}\label{eq:tmerge_general}
\frac{t_{\rm merge}^{\rm model}}{t_{\rm dyn}} = \frac{f(\epsilon)}{\ln [1+ \left( M_{\rm host}/M_{\rm sat} \right) ]} \left(\frac{M_{\rm host}}{M_{\rm sat}}\right)^\alpha  \left(\frac{R_{\rm c} (E)}{R_{\rm vir}}\right)^\beta \, .
\end{equation} 
Here $\epsilon$, $M_{\rm host}$, $M_{\rm sat}$, $R_{\rm c}(E)$, and $R_{\rm vir}$ are the circularity of the satellite's orbit, the masses of host and satellite halos, the circular radius of the satellite's orbit, and virial radius of the host, respectively (see \fref{merger_time_calculation}).

\begin{figure}[tpb]
\plotone{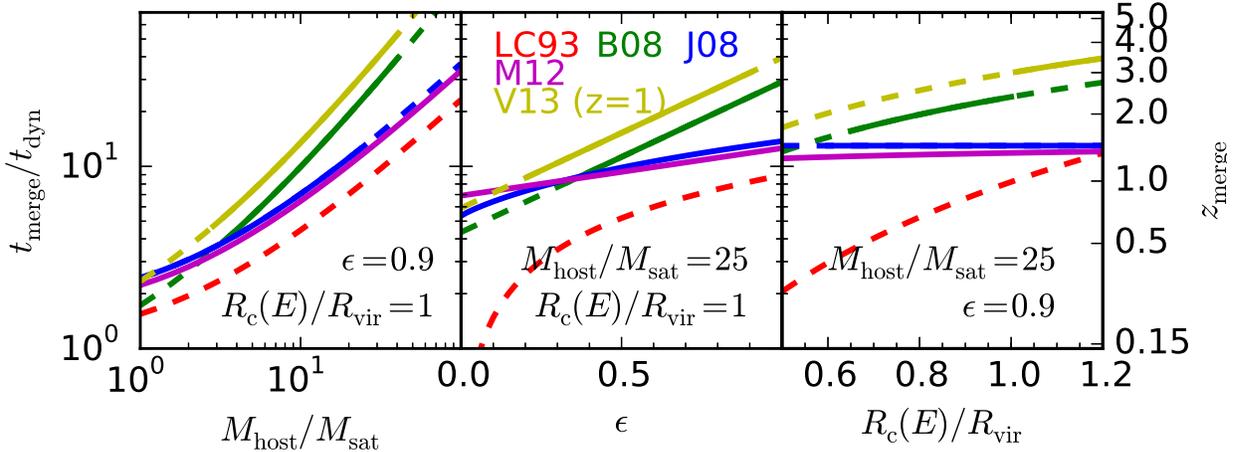}
\caption{Merger timescale as a function of the host-to-satellite mass ratio (left), the circularity of a satellite's orbit (middle), and the orbital energy (right).
In the right panel, we add ticks for the most recent merging epoch ($z_{\rm merge}$) of a satellite surviving until now.
Solid lines show a range of parameter space covered by numerical simulations from each literature.
The \V13 at $z = 0$ is not shown here because it is identical to the \B08.}\label{fig:merger_time_models}
\end{figure}

\fref{merger_time_models} shows the merger timescales of different models as a function of the host-to-satellite mass ratio, the circularity, and the orbital energy of the satellite's orbit.
The merger timescale monotonically increases with the parameters for the following reasons:
(1) A large host halo (i.e., large $M_{\rm host}$) has a long free-fall time.
(2) A compact satellite galaxy (i.e., small $M_{\rm sat}$) may suffer less tidal disruption.
(3) A satellite in a circular orbit (i.e., large $\epsilon$) would take more time to reach its host center than satellites have elongated orbits (i.e., small $\epsilon$).
(4) A satellite in a faster orbital motion (i.e., large $(R_{\rm c} (E) / R_{\rm vir})$) could survive longer.
 
The \LC93 has the shortest merger timescale among the models, which implies that the dynamical friction in simulations is usually lower than the analytic predictions from the ideal isothermal case.
Models derived from cosmological simulations (\J08 \& \M12) produce merger timescales that are similar to each other, and the same is true for those from isolated simulations (\B08 \& \V13).
For a major merger event ($M_{\rm host}/M_{\rm sat} \lesssim 3$), the merger timescale from cosmological simulations is slightly shorter than that from isolated simulations.
On the other hand, for a minor merger event ($M_{\rm host}/M_{\rm sat} \gtrsim 10$), $t_{\rm merge}$ from isolated simulations is always longer than that from cosmological simulations.
It may be partly because of the different setups between isolated and cosmological simulations, where the former simulates only a single merger event between two halos, while the latter includes multiple mergers.

The merger timescale from both isolated and cosmological simulations for a minor merger is longer than $\sim 5\,t_{\rm dyn}$.
As a result, in models from both types of simulations (\B08--\V13), satellite MBPs that suffered minor mergers after $z \simeq 1$ survive until $z = 0$ (see \fref{merger_time_models}).

\section{Results}\label{sec:results}

\subsection{Survival Probability of Satellite Galaxies}

\begin{figure}[tpb]
\centering
\vspace{-0.8cm}
\includegraphics[height=0.9\textheight]{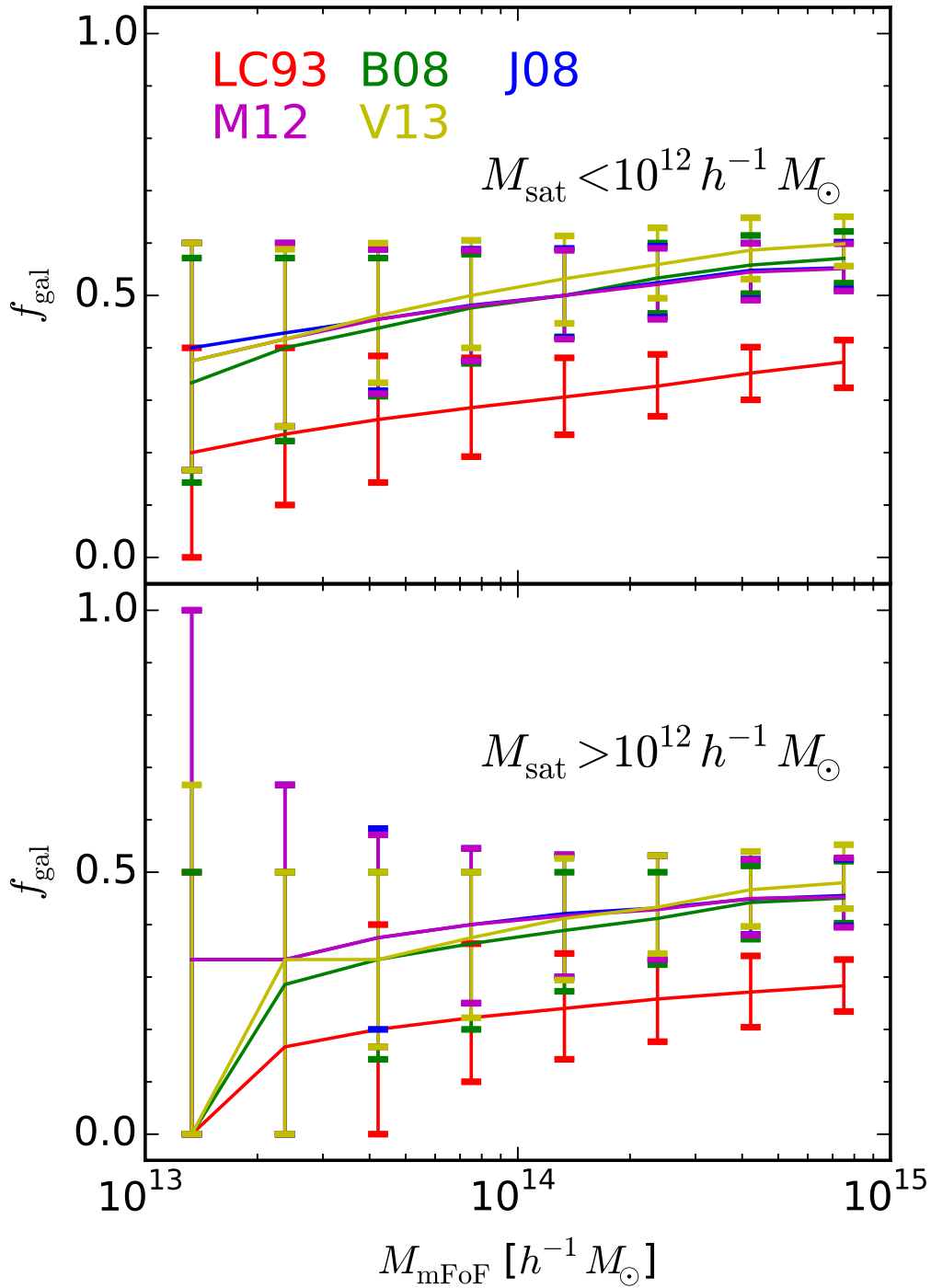}
\caption{
Distribution of the fraction of satellite MBPs of a halo that have survived until $z = 0$ among those who have merged into the halo during the simulation ($f_{\rm gal}$), as a function of the host halo mass ($M_{\rm mFoF}$).
Top: low-mass satellite population ($M_{\rm sat} < 10^{12} \Msunh$).
Bottom: massive satellite populations ($M_{\rm sat} > 10^{12} \Msunh$).
Error bars show $1\sigma$.}\label{fig:mHalo_fgal_massrange}
\end{figure}

\fref{mHalo_fgal_massrange} shows the distribution of the fraction of satellite MBPs of a halo that have survived until $z =0$ among those who have merged into the halo during the simulation (hereafter galaxy survival probability $f_{\rm gal}$).
Satellite MBPs are divided into two mass groups according to the infalling mass: massive satellites ($M_{\rm sat} > 10^{12} \Msunh$) and low-mass satellites ($M_{\rm mFoF}^{\rm min} < M_{\rm sat} < 10^{12} \Msunh$).

Since the merger timescale monotonically increases with the host-to-satellite mass ratio, $f_{\rm gal}$ monotonically increases with the host halo mass and decreases with the satellite mass.
In all cases, $f_{\rm gal}$ of satellites in massive hosts ($M_{\rm mFoF} \simeq 10^{15} \Msunh$) is about 20\% higher than that for low-mass hosts ($M_{\rm mFoF} \simeq 10^{13} \Msunh$).
Also, $f_{\rm gal}$ of low-mass satellites is about 10\% higher than that of massive satellites.

Because the \LC93 has the shortest merger timescale in most cases, $f_{\rm gal}$ in the \LC93 is about 20\% lower than the other models.
On the other hand, $f_{\rm gal}$ in \B08--\V13 agree quite well with one another.
This shows that the difference between isolated and cosmological simulations on the merger timescale, especially for minor mergers, does not significantly affect the overall satellite galaxy population (see \sref{models}).

\subsection{Galaxy Group Properties}\label{sec:group}

In this section, we study the physical properties of the simulated galaxy groups at $z=0$.
From now on, we equally divide the whole \HR4 simulation volume into 1000 cubic regions with a volume of $V=(315 \Mpch)^3$.
We simulate the distribution of galaxies in redshift space by adding the radial component of peculiar velocity to the radial coordinate of our mock galaxies.
In practice, we adopt the distant observer approximation and perturb mock galaxies along three Cartesian coordinate axes:
for example, the redshift space coordinate of galaxies observed along the $y$-axis is given by $\vec{x}_{\rm redshift} = \vec{x}_{\rm real} + \vec{\hat{y}} v_y / H_0$,
where $H_0 \equiv 100\,h \kmsMpc$, which is accurate at low redshifts.

As a comparison, we use the observed galaxy group catalog compiled from the volume-limited SDSS DR10 galaxies \citep{tempel2014}.
We use the volume-limited sample with an $r$-band absolute magnitude of $\mathcal{M}_r - 5 \log h < -20$, whose volume is about 1.27 times smaller than a single cubic region in our simulation.
For a fair comparison, we use a model of galaxy groups and their masses as described in \cite{tempel2014}, rather than the simulated halos and their true masses.
First, a galaxy group is modeled as a set of galaxies extracted in redshift space using the FoF method.
The FoF linking lengths in the radial and tangential directions are determined to satisfy the following conditions: 
(1) The radial-to-tangential linking length ratio is fixed as 10.
(2) The number of galaxy groups is maximized in a given galaxy sample.
In the volume-limited sample with an $r$-band absolute magnitude of $\mathcal{M}_r - 5 \log h < -20$, linking lengths in the tangential and radial directions are found to be $0.515$ and $5.15 \Mpch$, respectively \citep[see][]{tempel2014}.
Then we model the mass of a galaxy group with its radial velocity dispersion $\sigma_v$ and projected radius $R_{\perp}$, 
\begin{align}
\sigma_v^2 &\equiv \frac{1}{N_{\rm gal}-1} \sum_k (v_k - \langle v \rangle)^2, \\
R_{\perp}^2 &\equiv \frac{1}{N_{\rm gal}} \sum_k R_k^2,
\end{align}
by assuming that the group is virialized and that it follows the Navarro-Frenk-White (NFW) density profile (\citealt{navarro1997}; see \citealt{tempel2014} for details).
Here $N_{\rm gal}$ is the number of member galaxies in the group, and $\langle v \rangle$ is the average radial velocity of the group.
Hereafter, we call the modeled group mass the NFW mass ($M_{\rm NFW}$).

\begin{figure}[tpb]
\centering
\vspace{-0.8cm}
\includegraphics[height=0.9\textheight]{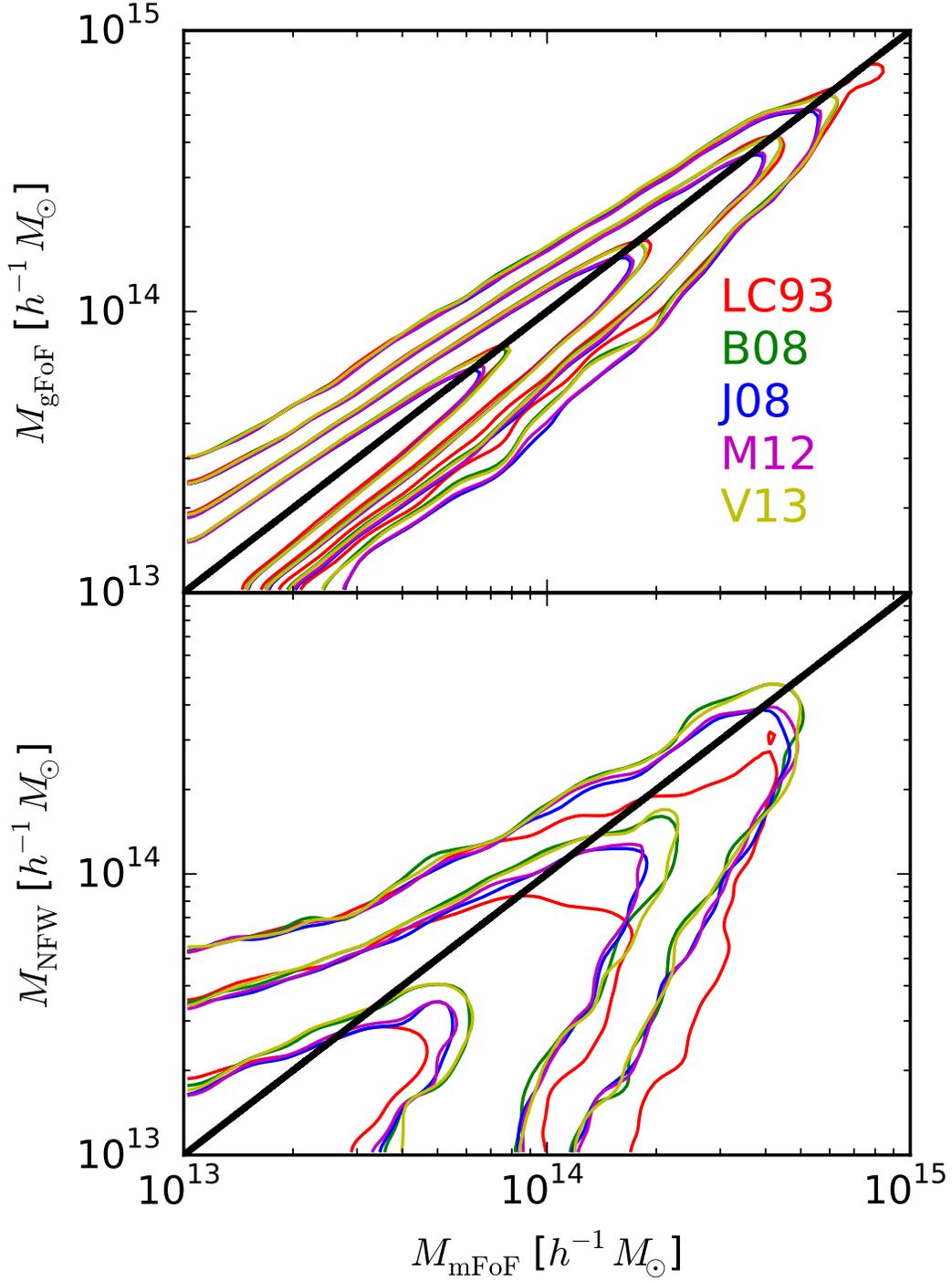}
\caption{Joint probability distribution functions of the FoF halo mass ($M_{\rm mFoF}$) and two modeled masses of its corresponding FoF galaxy groups in the mock galaxy samples.
Top: sum of MBP mass within a galaxy group ($M_{\rm gFoF}$).
Bottom: modeled group mass described in \cite{tempel2014} ($M_{\rm NFW}$).
See the text for details.
The contour level changes from $10^{-5.5}$ to $10^{-3.5}$ increasing by a factor of $10^{0.5}$.
}\label{fig:mHalo_mFoF_mNFW}
\end{figure}

\fref{mHalo_mFoF_mNFW} shows the correlation between the NFW mass and the corresponding FoF halo mass ($M_{\rm mFoF}$) in the \HR4 simulation.
Here, we link each FoF halo to a galaxy group that contains the central MBP of the halo.
While \cite{tempel2014} commented that the NFW mass estimation might be unreliable for poor galaxy groups ($M_{\rm NFW} \lesssim 10^{11} \Msunh$), we found that the NFW mass is also lower than the halo mass ($M_{\rm mFoF}$) for more than 80\% of rich groups ($M_{\rm NFW} \gtrsim 10^{13} \Msunh$).
Also, the correlation between the halo mass and the NFW mass is weak, as the Pearson's correlation coefficient between them is around $0.5$ (bottom).
To check whether the above disagreement comes from the difference of member galaxies between a galaxy group and its corresponding halo, we test another model of galaxy group mass defined as a sum of the MBP mass of all member galaxies ($M_{\rm gFoF}$).
Unlike the NFW mass, $M_{\rm gFoF}$ strongly correlates the halo mass with Pearson's correlation coefficient around 0.97 (top). 
This means that the difference between member galaxies does not play a significant role in the underestimation of NFW mass.

\begin{figure}[tpb]
\centering
\plottwo{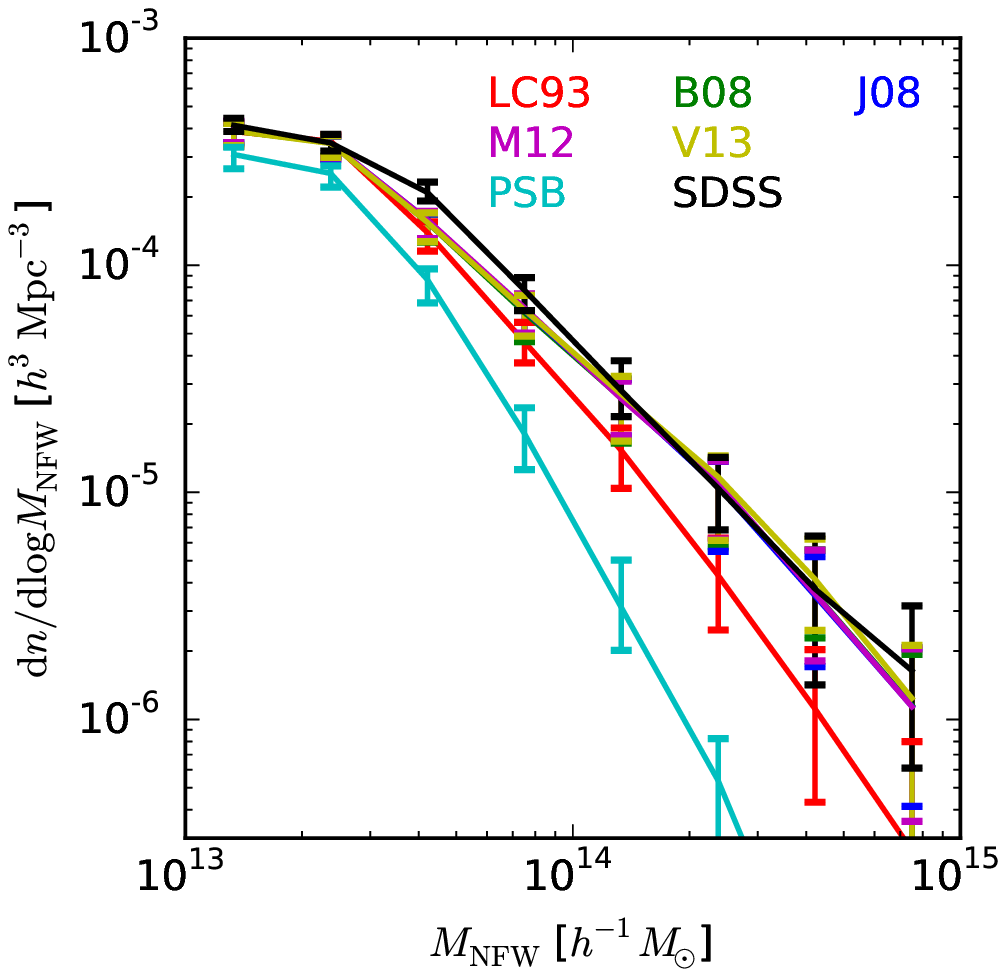}{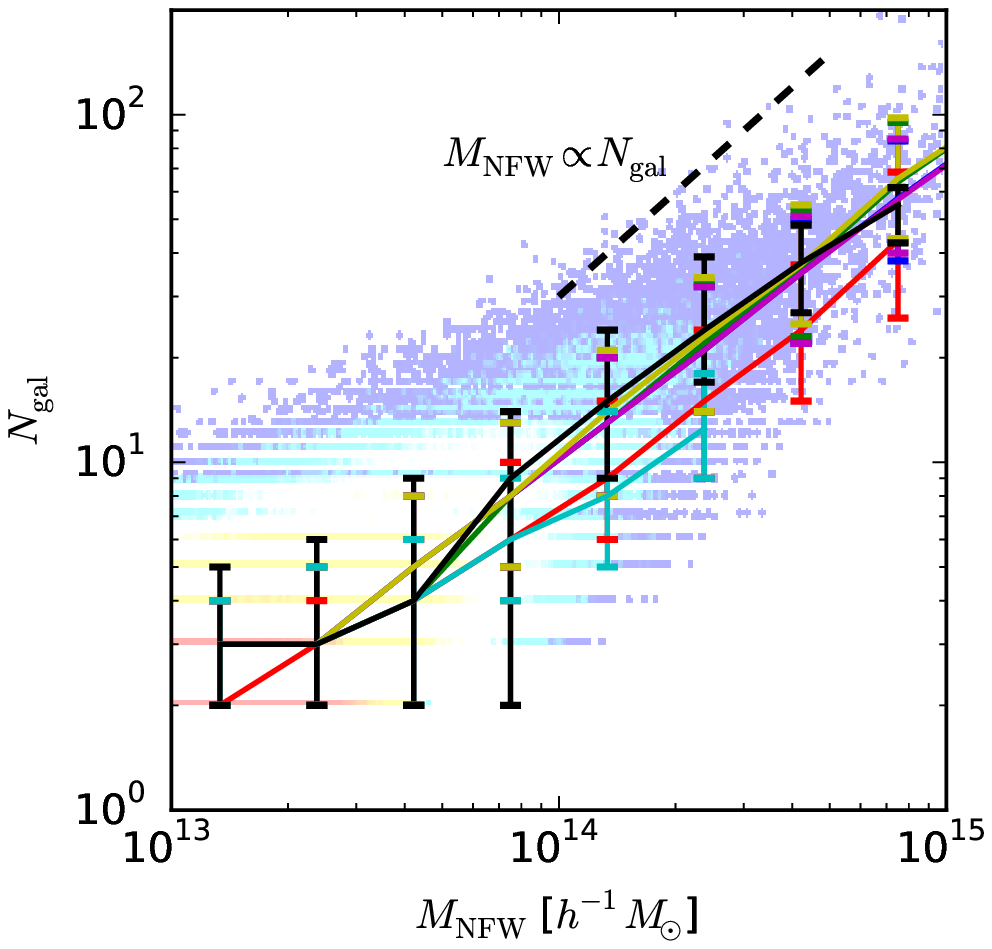}
\plottwo{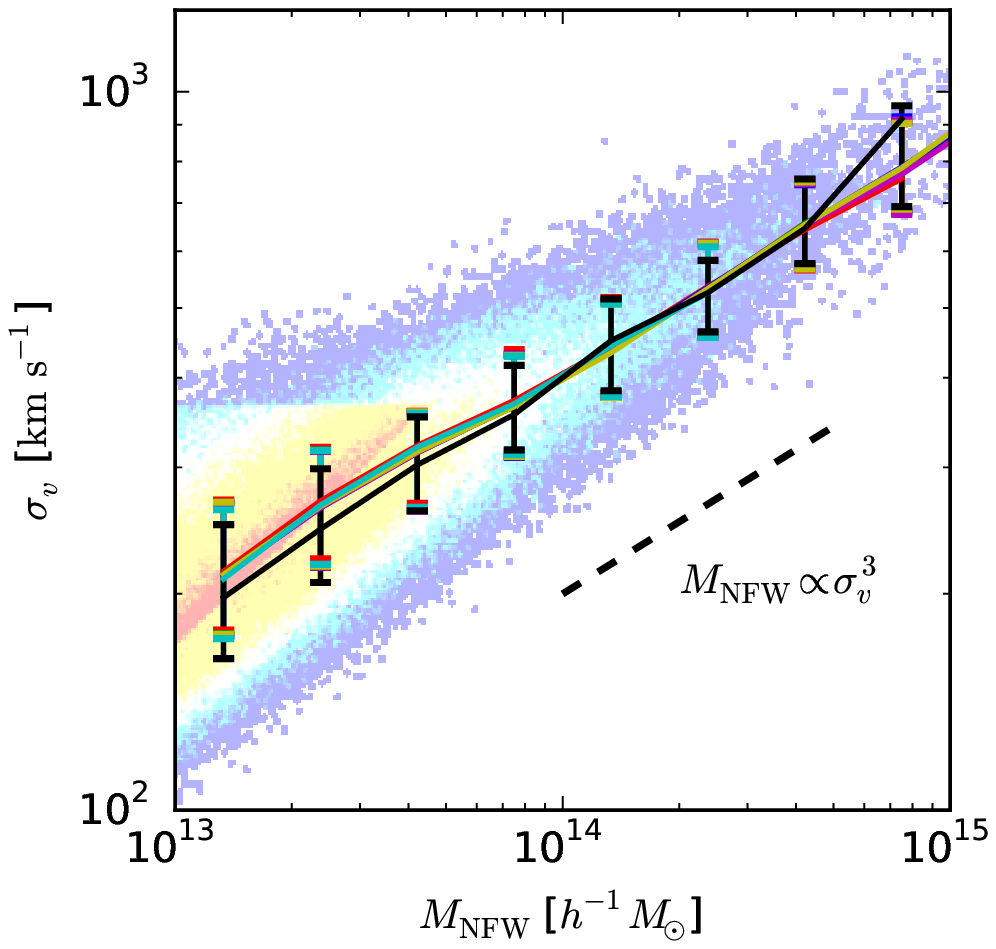}{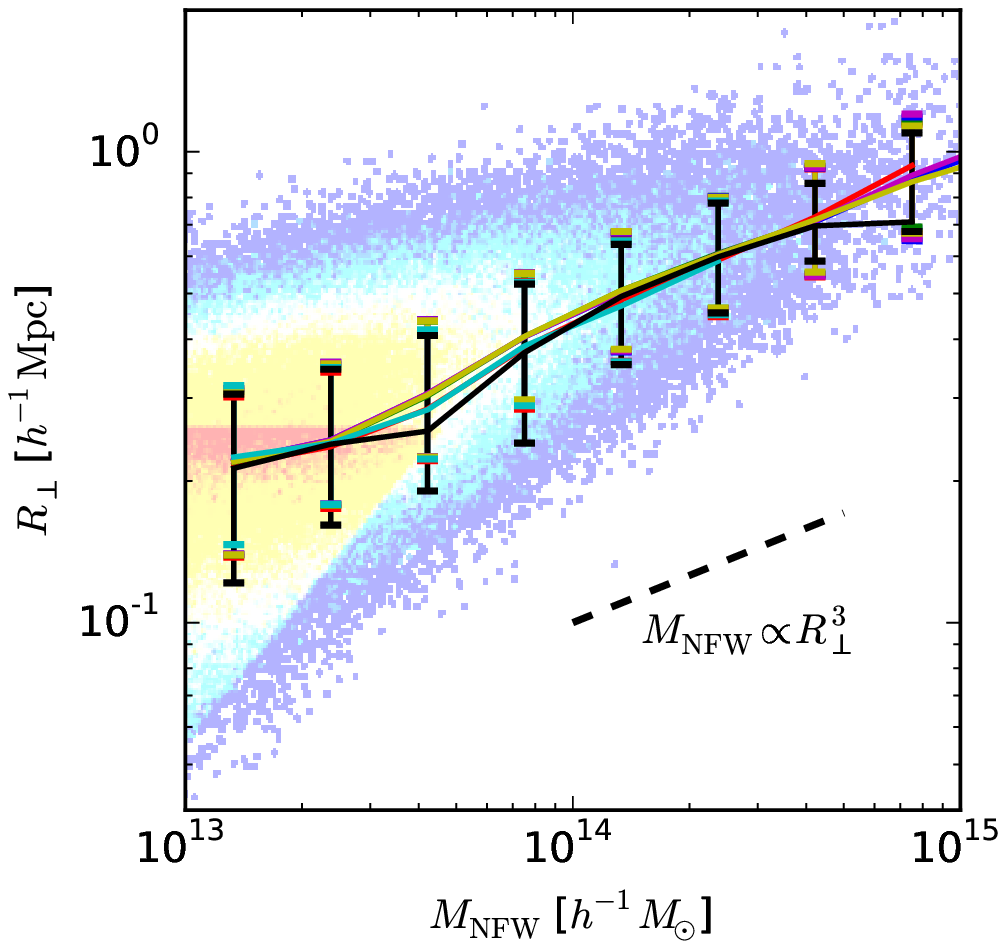}
\caption{Galaxy group properties of the mock galaxies (colored) and the volume-limited SDSS DR10 galaxies \citep[][black]{tempel2014} with an $r$-band absolute magnitude of $\mathcal{M}_r - 5 \log h < -20$.
Top left: galaxy group mass function in terms of NFW mass ($M_{\rm NFW}$).
Error bars show $1\sigma$ from the bootstrap resampling (SDSS) and 1000 equally divided regions (simulation).
Top right: the number of member galaxies ($N_{\rm gal}$) as a function of the NFW mass.
Error bars show $1\sigma$ in the galaxy group samples.
The dashed line shows the slope expected from the virial theorem.
The color map in the background displays the joint probability distribution in the \J08, coded on a logarithmic scale from low (blue) to high (red) probability.
Bottom left and bottom right: same as in the top right, but the radial velocity dispersion ($\sigma_v$) and the projected radius ($R_{\perp}$), respectively.}\label{fig:group}
\end{figure}

\fref{group} shows the mass function, the number of member galaxies, the radial velocity dispersion, and the projected radius of galaxy groups with $N_{\rm gal} \geq 2$, as a function of the NFW mass.
As well as the SDSS observation, all mock galaxy groups from the MBP-galaxy correspondence scheme (\LC93--\V13) and the PSB-galaxy correspondence scheme (hereafter \PSB) satisfy the virial theorem
\begin{equation}
M_{\rm NFW} \propto N_{\rm gal} \propto \sigma_v^3 \propto R_{\perp}^3 \, .
\end{equation}
The group properties in \B08--\V13 agree with the SDSS observation within $1\sigma$.
On the contrary, the \LC93 and the \PSB{} underestimate both the mass function and the number of member galaxies by more than $1\sigma$.
Nonetheless, the \LC93 and the \PSB{} have similar $M_{\rm NFW}$--$\sigma_v$ and $M_{\rm NFW}$--$R_{\perp}$ relations to the observation, mainly because the NFW mass fully depends on $\sigma_v$ and $R_{\perp}$.

\subsection{Two-point Correlation Function}\label{sec:2pcf}

In this section, we study the 2pCF and the projected 2pCF of the \HR4 mock galaxies measured in the tangential ($r_{\rm p}$) and radial ($\pi$) directions.
Numerous estimators of 2pCF have been suggested, but only subtle differences have been found between estimators \citep{davis1983, hamilton1993, landy1993}.
In this paper, we apply the \cite{hamilton1993} estimator,
\begin{equation}
\xi(r_{\rm p}, \pi) \equiv \frac{{\rm DD}(r_{\rm p}, \pi) {\rm RR}(r_{\rm p}, \pi)}{[{\rm DR}(r_{\rm p}, \pi)]^2} - 1 \, ,
\end{equation}
where DD, DR, and RR are the counts of data--data, data--random, and random--random pairs for a given $r_{\rm p}$ and $\pi$.
The projected 2pCF $w_{\rm p} (r_{\rm p})$ is
\begin{equation}
w_{\rm p} (r_{\rm p}) \equiv 2 \int_0^{\infty} d{\pi} \xi(r_{\rm p}, \pi)  \, .
\end{equation}
In practice, we perform the integration up to $\pi_{\rm max} = 40\Mpch$, and it is found that the choice of a larger value of $\pi_{\rm max}$ does not significantly affect our result.

We compare the 2pCFs of simulated galaxies with those of the volume-limited SDSS DR7 main galaxies \citep{zehavi2011}.
In addition to the previous $r$-band absolute magnitude condition $\mathcal{M}_r - 5 \log h < -20$, we use a brighter magnitude threshold $\mathcal{M}_r - 5 \log h < -21$, which leads to a galaxy number density of $\bar{n} \simeq 1.11 \times 10^{-3}\,h^3 {\rm Mpc^{-3}}$.

\begin{figure}[tpb]
\centering
\plottwo{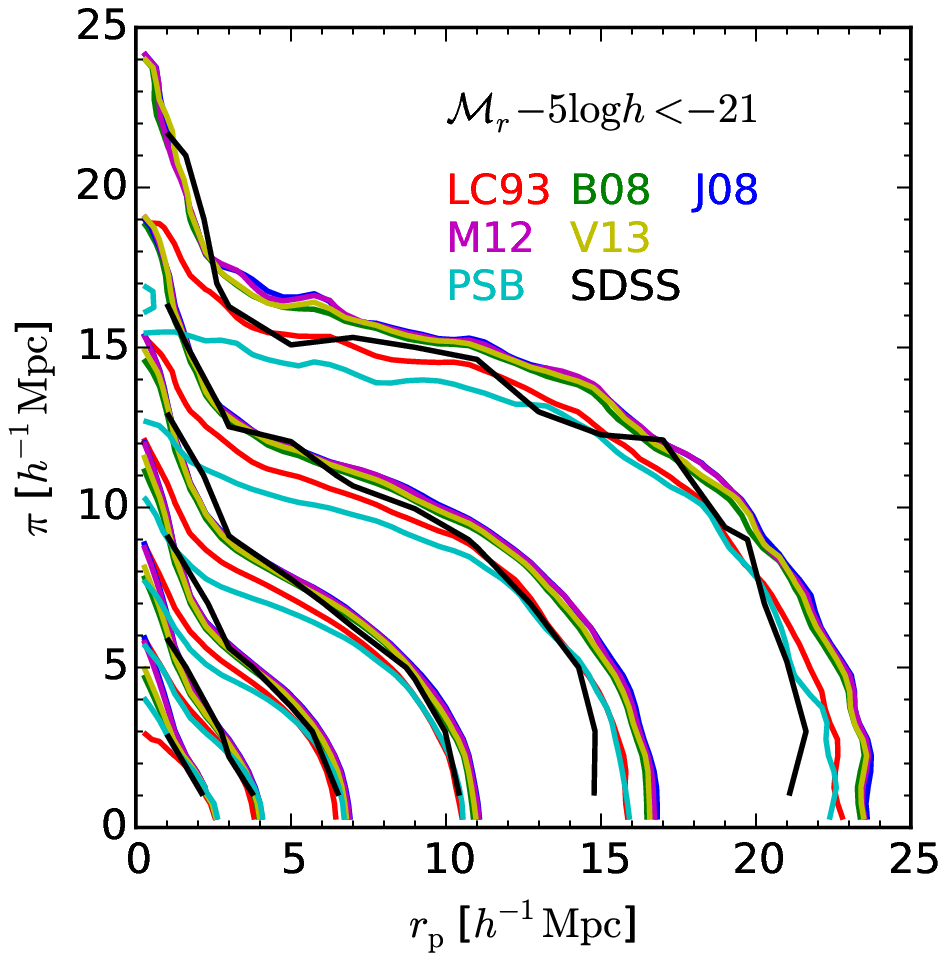}{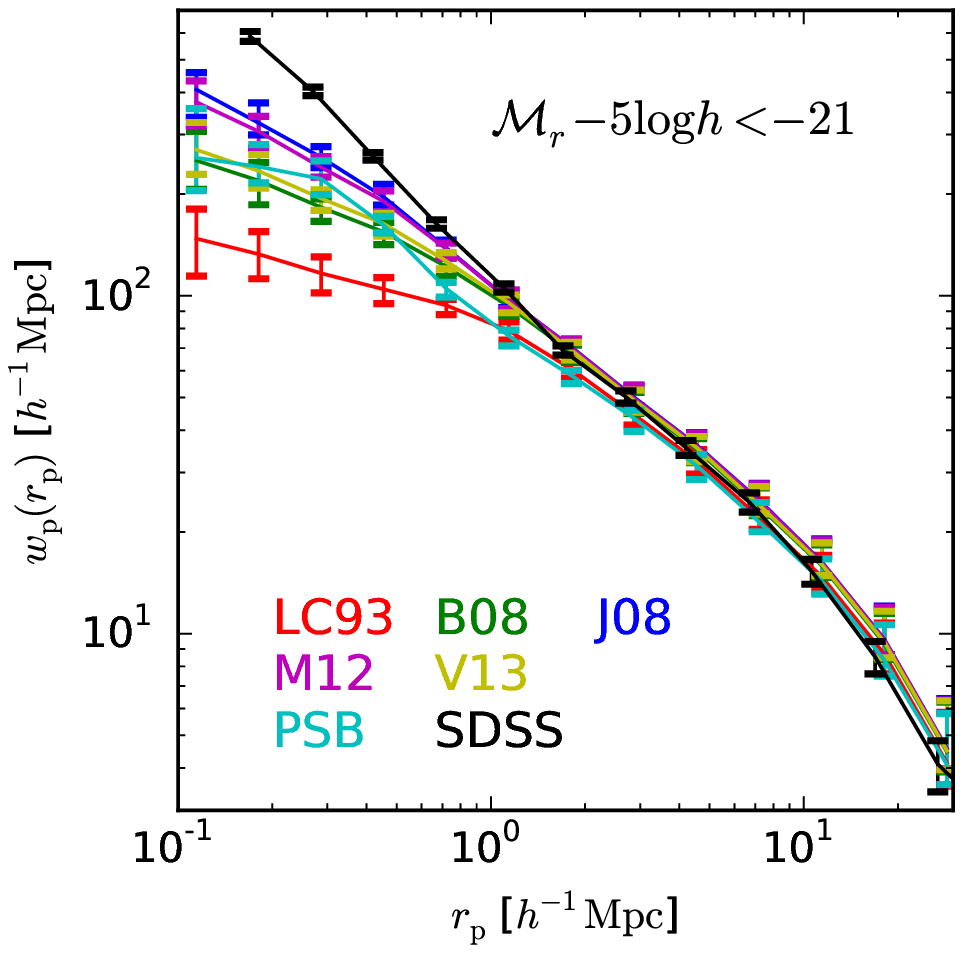}
\plottwo{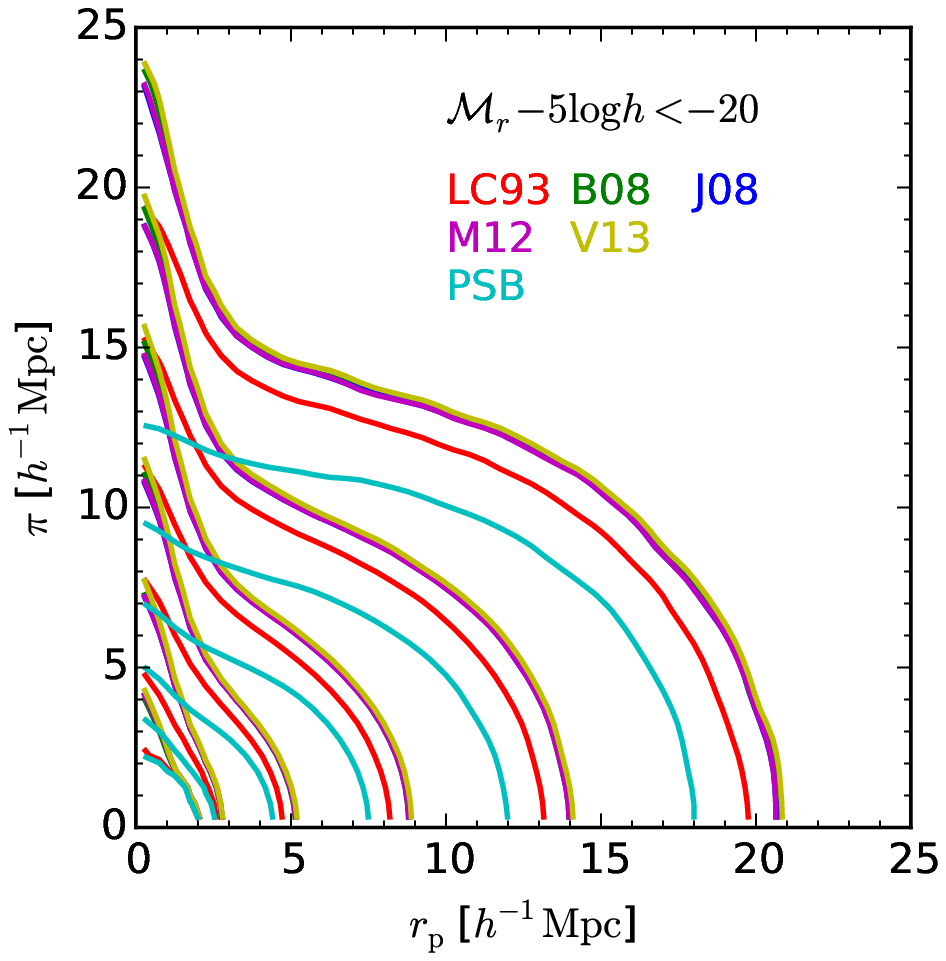}{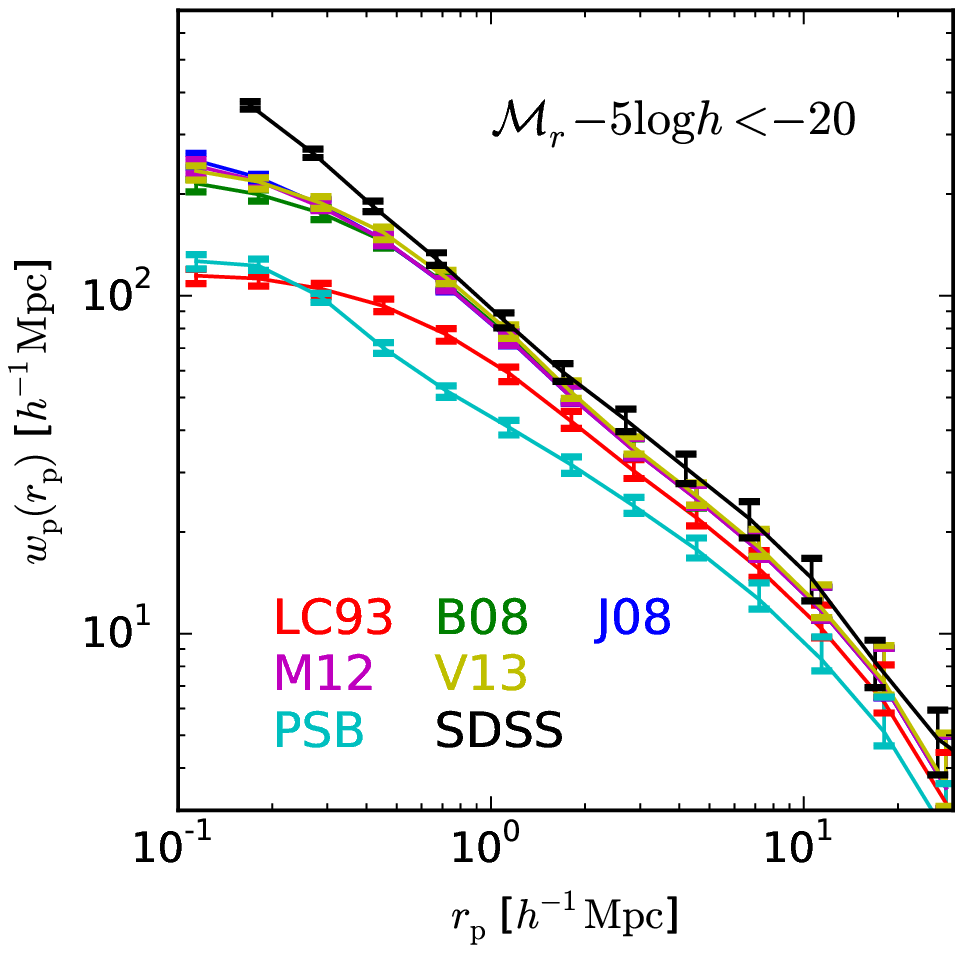}
\caption{Two-point correlation functions (2pCFs) of the mock galaxies (colored) and the volume-limited SDSS DR7 galaxies \citep[][black]{zehavi2011} with $r$-band absolute magnitudes of $\mathcal{M}_r - 5 \log h < -21$ (top) and $-20$ (bottom).
Left: contours of the 2pCFs $\xi(r_{\rm p}, \pi)$.
The contour level changes from 0.1875 to 6, increasing by a factor of 2.
Right: the projected 2pCFs $w_{\rm p}(r_{\rm p})$.
The error bars show $1\sigma$ from the jackknife resampling (SDSS) and 1000 equally divided regions (simulation).
The bottom left does not contain a result from the SDSS observation.}\label{fig:2pcf}
\end{figure}

\fref{2pcf} shows the 2pCFs and the projected 2pCFs of galaxy samples from observation and simulations on scales less than $30 \Mpch$.
All mock galaxy samples underestimate the 2pCFs on scales of $r_{\rm p} \lesssim 0.5 \Mpch$ because of the lack of the initial small-scale matter power below $\lambda^{\rm Nyquist} = 2\pi / k^{\rm Nyquist} = 1\Mpch$ in the \HR4.
On larger scales $r_{\rm p} \gtrsim 0.5 \Mpch$, the 2pCFs from \B08--\V13 agree with the observation within $1\sigma$.

Both the \LC93 and the \PSB{} reproduce the observed 2pCFs of brighter galaxies on large scales ($\mathcal{M}_r - 5 \log h < -21; r_{\rm p} \gtrsim 10 \Mpch$).
However, the \LC93 and the \PSB{} underestimate the 2pCFs of fainter galaxies and/or on small scales by more than $1\sigma$ and $2\sigma$, respectively.
Also, the 2pCF contour map in the \PSB{} does not show a Finger-of-God (FoG) feature.

\subsection{Difference between \LC93 \& \PSB{} and \B08--\V13}

In the previous sections, we have shown that the \LC93 and the \PSB{} underestimate the population of FoF galaxy groups, the number of member galaxies in a group, and the 2pCF, while \B08--\V13 reproduce the SDSS observations quite well.
Here we discuss what makes the \LC93 and the \PSB{}  substantially different from the rest.

As shown in \sref{models}, the \LC93 always produces shorter merger timescales than \B08--\V13.
It is because the \LC93 overestimates the dynamical friction by assuming that host halos follow an isothermal density profile.
Therefore, satellites close to the halo center are not properly identified in the \LC93 and satellites tend to be more distributed in the outer halo region.
On the other hand, to build a volume-limited mock galaxy sample with a fixed number density, one has to lower the mass limit in the \LC93.
Then satellites with small mass in the outer halo region become visible by lowering the mass limit.

The distance between satellites and their central galaxy in the \LC93 is longer than that in \B08--\V13, which leads to the following results.
(1) The number of detected FoF galaxy groups in the \LC93 is lower than that in other merger timescale models.
(2) The number of member galaxies in a given group in the \LC93 is lower than that in other groups (see \fref{group}).
(3) The 2pCF on small scales in the \LC93 is smaller than that in others.
The suppression of 2pCF is stronger for fainter galaxies since they tend to include more faint satellites (see \fref{2pcf}).

Similar to the \LC93, the \PSB{} lacks satellite galaxies close to their host center, mainly for the following reasons.
(1) The spatial resolution of the \HR4 ($d_{\rm mean} = 0.5 \Mpch$) is relatively low for finding subhalos close to the halo center.
(2) A subhalo-galaxy correspondence scheme cannot identify orphan galaxies by itself.
As a result, the \PSB{} also underestimates the FoF galaxy group population, the number of member galaxies in a galaxy group, and the 2pCF on small scales.
Moreover, the \PSB{} uses the bulk velocity of a subhalo for the galaxy velocity, which tends to produce a smaller peculiar velocity than our MBP-galaxy correspondence scheme. 
Therefore, the \PSB{} lacks the FoG feature in the 2pCF contour map (see \fref{2pcf}).

\section{Summary}\label{sec:summary}

In this paper, we introduced an MBP-galaxy correspondence scheme that applies the modeled merger timescales to the fate of MBPs.
We adopted five models for the merger timescale:
one from analytic calculation \citep{lacey1993}, 
two from isolated halo simulations \citep{boylan-kolchin2008, villalobos2013}, 
and two from cosmological simulations \citep{jiang2008, mccavana2012}.

To produce the mock galaxy samples, we applied our MBP-galaxy correspondence scheme to the Horizon Run 4 simulation, which covers the comoving volume of $(3150 \Mpch)^3$ with $6300^3$ particles \citep{kim2015}.
In addition to the five sets of mock galaxy samples derived from the above five models, we produced a mock galaxy sample by applying a typical subhalo-galaxy correspondence scheme \citep{kim2008}.
We compared several properties of galaxy groups and the 2pCFs of our mock galaxies with the volume-limited SDSS galaxies with the $r$-band absolute magnitudes of $\mathcal{M}_r - 5 \log h < -21$ and $-20$ \citep{zehavi2011,tempel2014}.

Because of our limited simulation resolution, the subhalo-galaxy correspondence scheme underestimates the population of satellite galaxies close to their host center. 
Also, because the subhalo-galaxy correspondence scheme uses the bulk velocity of a subhalo for the galaxy peculiar velocity, it suppresses the FoG feature. 
As a result, the subhalo-galaxy correspondence scheme underestimates the populations of galaxy groups and the 2pCFs by more than $2\sigma$.

On the contrary, the MBP-galaxy correspondence scheme with models based on numerical simulations reproduces the observed galaxy-group properties and the 2pCFs within $1\sigma$. 
While the merger timescale of minor mergers in isolated and cosmological simulations do not agree with each other, it is found that this lack of agreement does not significantly affect the overall population of satellite galaxies at $z=0$.

We also tested the relations among the group mass, the radial velocity dispersion, and the projected size.
All mock galaxies, including those from a subhalo-galaxy correspondence scheme, reproduce the observed relations.
However, such an agreement might be rather artificial because the adopted model of group mass fully depends on the radial velocity dispersion and the projected size \citep[see][]{tempel2014}.
For a further study on the validity of galaxy group properties, one may need to use a model of group mass that does not depend on the velocity dispersion and the projected size.

\acknowledgments{
The authors thank the Korea Institute for Advanced Study for providing computing resources (KIAS Center for Advanced Computation Linux Cluster System) for this work.

This work was supported by the Supercomputing Center/Korea Institute of Science and Technology Information, with supercomputing resources including technical support (KSC-2013-G2-003) and the simulation data were transferred through a high-speed network provided by KREONET/GLORIAD.

The authors thank Idit Zehavi for providing us with the two-point correlation function data from the SDSS galaxy sample.
The authors thank Ho Seong Hwang and an anonymous referee for their constructive suggestions that leads to the improvement of this paper.

Funding for the SDSS and SDSS-II has been provided by the Alfred P. Sloan Foundation, the Participating Institutions, the National Science Foundation, the US Department of Energy, the National Aeronautics and Space Administration, the Japanese Monbukagakusho, the Max Planck Society, and the Higher Education Funding Council for England. The SDSS Web site is http://www.sdss.org/.

The SDSS is managed by the Astrophysical Research Consortium for the Participating Institutions. The Participating Institutions are the American Museum of Natural History, Astrophysical Institute Potsdam, University of Basel, University of Cambridge, Case Western Reserve University, University of Chicago, Drexel University, Fermilab, the Institute for Advanced Study, the Japan Participation Group, Johns Hopkins University, the Joint Institute for Nuclear Astrophysics, the Kavli Institute for Particle Astrophysics and Cosmology, the Korean Scientist Group, the Chinese Academy of Sciences (LAMOST), Los Alamos National Laboratory, the Max Planck Institute for Astronomy (MPIA), the Max Planck Institute for Astrophysics (MPA), New Mexico State University, Ohio State University, University of Pittsburgh, University of Portsmouth, Princeton University, the US Naval Observatory, and the University of Washington.
}

\end{document}